\documentclass[preprint]{revtex4-1}
\usepackage{graphicx}
\usepackage{color}
\newcommand{\fig}[4]{\begin{figure}
\centering
\includegraphics[width=#2\textwidth]{#1}
\caption{#3}
\label{#4}
\end{figure}}
\begin{document}
\title{Probing near-interface ferroelectricity by conductance modulation of a nano-granular metal}
\author{M.\ Huth}
\author{A.\ Rippert}
\author{R.\ Sachser}
\author{L.\ Keller}
\affiliation{Physikalisches Institut, Goethe University, Frankfurt am Main, Germany}
\begin{abstract}
The electronic functionality of thin films is governed by their interfaces. This is very important for the ferroelectric (FE) state which depends on thin-film clamping and interfacial charge transfer. Here we show that in a heterostructure consisting of a nano-granular metal and an organic FE layer of [tetrathiafulvalene]$^{+\delta}$[p-chloranil]$^{-\delta}$ the nano-granular layer's conductance provides a sensitive and non-invasive probe of the temperature-dependent dielectric properties of the FE layer. We provide a theoretical framework that is able to qualitatively reproduce the observed conductance changes taking the anisotropy of the dielectric anomaly at the paraelectric(PE)-FE phase transition into account. The approach is also suitable for observing dynamical effects close to the phase transition. Focused electron beam induced deposition as fabrication method for the nano-granular metal guarantees excellent down-scaling capabilities, so that monitoring the FE state on the lateral scale down to 20--30\,nm can be envisioned.
\end{abstract}
\maketitle
\section{Introduction}
It is thin films and, in particular, their interfaces that define the electronic functionality of devices. This is an established fact in inorganic semiconductors as has been demonstrated for oxides in recent years \cite{Mannhart2010a}. The ferroelectric (FE) state provides important functionality in this regard, since the ability to switch the electric polarization by an electric field is the basic property needed in ferroelectric random access memories or field effect transistors \cite{Dawber2005a}. The paraelectric (PE) to FE phase transition at the Curie temperature $T_C$ is equally important in basic research, since it is one of the most representative phase transitions in solids. One highly relevant issue is the character of the PE-FE transition when it is tuned to 0 K, i.e.\ towards a possible quantum critical point. In this case the phase transition is driven by quantum fluctuations instead of thermal fluctuations with important consequences for the electronic behavior even at elevated temperatures \cite{Horiuchi2003a}. As a second important issue we mention the fact that the electric polarization can only be correctly represented on the basis of the quantum mechanical Berry phase concept which allows for the calculation of the displacement current under changes of the electronic wave functions taking place as the PE-FE phase boundary is crossed \cite{Resta1994a}.

For both, applied and basic research, the ability to tune the PE-FE transition to a large degree is very attractive. How can this tuning be accomplished? Two mechanisms are particularly suitable for thin film structures, namely clamping effects and interfacial charge transfer. Clamping effects occur, if a thin film strongly adheres to a bulk substrate and follows its elastic response, e.g., thermal expansion effects in the consequence of temperature changes. Interfacial charge transfer is governed by the interfacial electronic states which form when the thin film is deposited on a substrate surface or as part of a multi-component heterostructure. Organic FE from the class of charge transfer systems (CTS) are ideal model systems in this regard, since their reaction to clamping effects and charge transfer over the interface is particularly strongly pronounced \cite{Kawasugi2008a, Kawasugi2009a, Solovyeva2009a, Solovyeva2011a}. This is so because their thermal expansion coefficients are about an order of magnitude larger than those of oxides \cite{Mueller2002a} and their FE properties are governed by intermolecular charge transfer between a donor and an acceptor molecule rather than simple displacement of static point charges on the molecules. This gives rise to an electronic FE polarization which is much stronger than it would be expected for a simple displacive type FE state \cite{Kobayashi2012a}. At the same time they are multiferroics, since with the onset of FE also a one-dimensional antiferromagnetic state is formed \cite{Filatov2013a}.

An important question in analyzing the FE state in organic CTS is: How can the influence of clamping and charge transfer be monitored, ideally in a way that does not interfere with the FE state itself? This is the starting point of the work presented here. We introduce a novel concept for monitoring the FE state in thin films with large sensitivity to the near-interface region. We show that strong modulations in the Coulomb charging energy in a nano-granular metal occur, if this is in close proximity to a FE thin film. These modulations lead to easily detectable changes in the electrical conductance which thus becomes a minimally invasive probe of the FE state, its temperature and electric field dependence, as well as its dynamic properties.

We demonstrate the monitoring effect for a bilayer thin film structure (see schematic in \ref{fig4}) consisting of the organic ferroelectric [tetrathiafulvalene]$^{+\delta}$[p-chloranil]$^{-\delta}$ (TTF-CA) and a Pt-based nano-granular metal (Pt(C)) prepared by focused electron beam induced deposition (FEBID). TTF-CA is an organic CTS that shows a valence instability, also called neutral-ionic phase transition, at 81\,K which drives the system from a PE to a FE state \cite{Torrance1981a}. It consists of one-dimensional, mixed stacks of the donor (D), TTF, and the acceptor (A), CA. The degree of charge transfer $\delta$ is about 0.2 from D to A in the PE phase and increases by a first order phase transition to about 0.7 in the FE state. With the PE-FE phase transition a dimerization of the D/A-pairs is associated (see also \ref{fig1}) leading to a multiferroic ground state in which, in addition to the FE state, a one-dimensional dimerized antiferromagnetic spin $S=1/2$-state with an exchange integral of the order of 1100\,cm$^{-1}$ is formed \cite{Filatov2013a, Giovannetti2009a, Mitani1984a}. Nano-granular Pt(C) consists of fcc-Pt crystallites of 1.5--5\,nm diameter embedded in an amorphous carbon matrix. Pt(C) grown by FEBID has shown to provide a model system for studying charge transport effects in nano-granular metals with tunable inter-grain coupling strength \cite{Huth2012a, Sachser2011a}. In a nano-granular metal charge transport occurs via inelastic tunneling between the metallic grains. Due to the very small size of the individual Pt crystallites, this charge transport is governed by the charging energy $E_C$ associated with each tunnel event. It is this charging energy which is very sensitive to the dielectric properties of neighboring dielectric layers. As a consequence, the onset of the FE state associated with a pronounced increase of the dielectric constant $\varepsilon_r(T)$ of TTF-CA can be detected as a very pronounced change in the temperature-dependent conductance of the nano-granular metal. We provide a theoretical framework to understand the observed conductance changes and provide exemplary model calculations that take the anisotropy of the dielectric anomaly at the PE/FE phase transition into account. Our data suggest that this approach is very suitable for observing dynamical effects close to the PE-FE phase transition, such as fluctuations in the dielectric constants due to the coexistence of PE and FE domains or the movement of FE domains. Because of the FEBID fabrication method of the nano-granular metal, our approach provides excellent down-scaling capabilities. Monitoring the FE state on the lateral scale down to the limit defined by FEBID, i.~e.\ 20--30\,nm for nanostructures with electrical contact leads, can therefore be envisioned.
\section{Results}
We begin by presenting our results for polycrystalline TTF-CA films of several micrometer thickness. By employing interdigitized electrode structures, and thus strongly increasing the effective thin film cross section, the conductance was sufficiently large to follow the onset of the FE state by the associated conductance anomaly of TTF-CA itself, as is shown in the main panel of Fig.~\ref{fig1}. We were able to follow the conductance down to about 45\,K below which the film resistance became larger than the isolation resistance of our setup. In parallel, we could qualitatively monitor the dielectric properties of TTF-CA by the change of the capacitance of the interdigitized electrode structure with TTF-CA coverage in the vicinity of the phase transition, as shown in the inset of Fig.~\ref{fig1}. As is evident, the phase transition occurs at 81\,K accompanied by a strong increase of the conductance and capacitance. This behavior is also typical for bulk crystals. However, for thin films to exhibit clamping effects layer thicknesses in the 100\,nm range and below are necessary. For such thin layers direct conductance measurements are extremely demanding or even impossible. This becomes apparent, if one compares the resistance values of about 440\,M$\Omega$ at the phase transition rapidly growing to several ten G$\Omega$ at lower temperatures even for the several micrometer thick TTF-CA thin film on an optimized electrode structure as shown in Fig.~\ref{fig1}a). Moreover, the additional charge carriers injected into the FE will disturb the charge carrier dynamics intrinsic to the FE itself. For TTF-CA, in particular, it is charged domain boundaries governed by a unique soliton dynamics which are mostly responsible for the charge transport properties \cite{Kagawa2010a}. In order to be able to study both, clamping effects and the intrinsic dynamics of the FE we introduce a different approach.

By careful growth studies we were able to find conditions for a controlled growth of polycrystalline TTF-CA thin films in the several 10\,nm-range. In Fig.~\ref{fig2} we present a typical atomic force microscopy (AFM) image of a nominally 50\,nm thick TTF-CA layer grown on an amorphous SiO$_2$ layer (see Methods section for details). The polycrystalline nature of these films is evident from specular X-ray diffraction (XRD) scans, as shown in the lower right panel of Fig.~\ref{fig2}. This same growth behavior is observed on top of 10\,nm to 100\,nm thick microstrips of nano-granular Pt(C) grown between two metallic electrodes by FEBID, as indicated by the dashed outlines in the AFM image. We note, that the Pt(C) microstrip is covered by about five to ten TTF-CA islands which will, in general, differ in their growth orientation. A SEM image of a typical microstrip taken directly after FEB-induced growth is also shown in Fig.~\ref{fig2}.

In the resulting double layer structure of Pt(C) and TTF-CA the application of a voltage between the electrodes leads to an electric field acting in parallel along the nano-granular metal and the organic FE. Due to the very large resistivity of the organic thin film the current flow is, however, restricted to the Pt(C) microstrip. The strong increase of the dielectric constant of TTF-CA at the PE-FE phase transition has now dramatic consequences on the charge transport in the nano-granular metal.

In Fig.~\ref{fig3} we show the conductance vs.\ temperature behavior of three Pt(C) microstrips with different thicknesses in the range from 10\,nm to 100\,nm. Each of the microstrips is covered with a 50\,nm thick polycrystalline layer of TTF-CA. For reference purposes we also
include conductance data obtained on analogous Pt(C) microstrips without TTF-CA coverage. At this point we make several important observations: (1) In contrast to the pronounced thermally activated behavior of the conductance of uncovered Pt(C) microstrips, for TTF-CA covered Pt(C) the reduction of the conductance as the temperature is reduced is much less pronounced. (2) The microstrips show a pronounced increase of the apparent conductance noise in the temperature range 50\,K to 65\,K (100\,nm sample) and 50\,K to 56\,K (20\,nm sample), respectively. For the 100\,nm thick microstrip the noisy data follows a peak-like feature. This becomes also apparent for the 20\,nm microstrip after applying a running average. In addition, the noise onset for the 20\,nm sample coincides with the peak observed for the 100\,nm sample. (3) For the 10\,nm microstrip the overall temperature dependence of the conductance, albeit still thermally activated, is very weak. No peak is observed in the measured temperature range.

From these observations we draw the following conclusions: (a) The peak-like anomalies between 50\,K and 65\,K are caused by the PE-FE transition in the TTF-CA top layer. These anomalies are the consequence of a strong reduction of the Coulomb charging energy in the nano-granular material and will be discussed in more detail in the following section. (b) The rather large dielectric constant $\varepsilon_r^{(a)}\approx 500$ of TTF-CA along the a-axis, even out of the immediate PE-FE phase transition region, also leads to a reduced Charging energy which results in a weakening of the thermally activated transport behavior. (c) Comparing the data for the 20\,nm and 100\,nm microstrip, a different shape of the dielectric anomaly in TTF-CA at the phase transition is observed. We associate this to the crystallographic orientation of the majority of the TTF-CA growth domains on top of the respective Pt(C) microstrips in conjunction with the large dielectric anisotropy of TTF-CA when comparing the $a$- and $c^*$-orientation. (d) For the 10\,nm microstrip the charging energy renormalization effect is most pronounced. In particular, we assume here a predominant $a$-axis orientation of the TTF-CA top layer which results in a strong suppression of the thermally activated conductance behavior. In addition, for typical values of $\varepsilon_r^{(a)}$ up to and larger than 500 at the PE-FE transition the charging energy renormalization becomes saturated over a wide temperature range which hinders the observability of a peak structure at the phase transition. In the next section we present a rationale for these conclusions.
\section{Discussion}
In the following we will first briefly compile the most relevant dielectric properties of TTF-CA bulk crystals for reference purposes. Next, some important recent results on the temperature-dependent conductivity of nano-granular metals will be presented. This provides the basis for the main part of this section which is devoted to the modeling of the influence of the dielectric properties of the substrate/nano-granular/PE-FE trilayer system. We conclude this section by a preliminary discussion of possible reasons for the observed reduction of the Curie temperature from 81\,K to 56\,K.

For bulk single crystals of TTF-CA the PE-FE phase transition sets in at 81\,K and influences the dielectric and electric properties over an extended temperature range down to about 50\,K. This is assumed to be the consequence of an extended coexistence region of the PE and FE phase \cite{Cointe2003a}. The dielectric response at the Curie point is highly anisotropic both, with $\varepsilon_r$ reaching values up to and above 500 for an electric field along the $a$-axis and about 50 along the $c^*$-axis at the phase transition. This is schematically indicated in the inset of the main panel of Fig.~\ref{fig4}. Due to the highly anisotropic electronic structure in TTF-CA the electrons are mostly confined to 1D mixed-stack chains. Consequently, the PE-FE phase transition has the character of a Peierls-like instability and is associated with a dimerization of the D/A-pairs, as is indicated in the right panel of Fig.~\ref{fig1}. In the ionic phase two energetically degenerate dimerization patterns coexist. This degeneracy occurs because the additional charge transfer at the PE-FE transition can occur equally likely towards the positive or negative stacking axis direction. The boundaries between the degenerate FE domains have soliton character and they occur as either spin-carrying or spinless solitons. Both have fractional charges but of opposite sign. For further details we refer to \cite{Kagawa2010a}. Here, it is important to note that the soliton dynamics is directly reflected in TTF-CA on the local scale and can thus cause a local, time-dependent modulation of the charging energy in the underlying nano-granular Pt(C) microstrip. This will be relevant for understanding the apparent conductance noise in the Pt(C) microstrips observed in proximity of the PE-FE phase transition.

Charge transport in nano-granular metals represents a very challenging problem within the topical regime of correlated electron systems. This is due to the complex interplay of Coulomb interaction, disorder and finite size effects. Only very recently, significant new theoretical insight could be gained in the limits of weak, $g\ll 1$, and strong, $g\gg 1$, inter-grain tunnel coupling strength (see \cite{Beloborodov2007a} for a recent review) which has been experimentally confirmed (see, e.g., \cite{Sachser2011a}). Here $g$ is given in units of the quantum conductance $G=2e^2/h$. Within the weak-coupling regime, of relevance here, the dominant transport mechanism
is based on a variable-range type hopping (VRH) which takes Coulomb interactions into account. In more detail, as an electron tunnel event between neighboring grains occurs an associated charging energy $E_C=e^2/2C$ ($C$: capacitance of a grain in the effective surrounding nano-granular medium) has to be overcome. As the temperature is reduced, however, correlated tunneling of more than one electron can occur and sequential inelastic co-tunneling becomes the dominating transport channel. This leads to the following conductivity vs.\ temperature behavior \cite{Beloborodov2007a}
\begin{equation}
\sigma = \sigma_0\exp{\left[-\left(T_0/T\right)^{1/2}\right]}
\end{equation}
with
\begin{equation}
k_BT_0 \approx e^2/4\pi\varepsilon_0\varepsilon_r\xi(T)
\end{equation}
whereby a weakly temperature-dependent attenuation length $\xi(T)$ of the electronic wave function is introduced. Here it is important to note that the activation temperature $T_0$ depends on two accounts on the dielectric constant experienced by the embedded metallic grain. First, $T_0$ is inversely proportional to the effective dielectric constant $\varepsilon_r$ of the matrix in which the grain is embedded. Second, $\xi$ depends also on the effective dielectric constant by way of its dependence on the charging energy \cite{Beloborodov2007a}
\begin{equation}
\xi(T) = 2D/\ln{\left(E_C^2/16\pi g(k_BT)^2\right)}
\end{equation}
where $D$ denotes the grain diameter. For a spherical grain the capacitance is $C=2\pi\varepsilon_0\varepsilon_r D$, so that $E_C$ is also inversely proportional to $\varepsilon_r$.

We now proceed to presenting a modeling approach that contains the essential aspects on a mean-field level for describing the observed conductance anomalies of the Pt(C) microstrips with TTF-CA top layer. The key point in the modeling approach lies in the modification of the effective dielectric constant experienced by the metallic grains at various distances to the Pt(C)/TTF-CA interface. We proceed in two steps. In the first step, the Coulomb potential at a given point $\mathbf{r}$ created by a point charge at position $\mathbf{r}'$ in the second layer of a trilayer heterostructure (see schematics in Fig.~\ref{fig4}) is calculated. This can be accomplished by an repeated image charge method leading to an expression for the Coulomb potential $e/4\pi\varepsilon_0\varepsilon_r(r)r$. In the second step, the capacitance of a sphere in the second layer can be obtained by the source point collocation method which uses the fact that the surface of a metallic sphere is an equipotential surface. On $n$ different positions, corresponding to the sphere surface, collocation points are defined. On these the same potential $V$ is enforced. The potential is created by a collection of $n$ point charges positioned inside the spatial region corresponding to the sphere volume. The appropriate charge values $\{Q_i\}$ of the point charges are obtained from the solution of a corresponding $n$-dimensional equation system. The capacitance is then obtained via $C=\sum_{i=1}^nQ_i/V$. For details of this procedure we refer to a different publication \cite{Huth2014a}.

In Fig.~\ref{fig4} (right) we show exemplary results for the charging energy $E_C(z)=e^2/2C(z)$ as a function of the distance from the Pt(C)/TTF-CA interface assuming different dielectric constants for the TTF-CA layer as indicated. Two trends are apparent. First, strong renormalization effects due to the proximity to the high-dielectric layer are confined to the near-interface region with a penetration depth of about 20\,nm. Second, for $\varepsilon_r^{TTF-CA}$ larger than about 100 the renormalization tends to saturate.

We now use the results for $C(z)=2\pi\varepsilon_0\varepsilon_r(z)D$ to calculate the expected qualitative temperature dependence of the conductance of the Pt(C) microstrips. For this we rely on a parallel-circuit model for the nano-granular metal. We assume that it consists of equidistant layers of regularly arranged spheres of equal diameter, as is schematically indicated in Fig.~\ref{fig4}. We use the respective $z$-dependent values of the capacitance and dielectric constant to obtain the respective layer conductances $\{\sigma(T;z_i)\}$ and obtain the overall conductance by summation: $\sigma(T) = \sum_i\sigma(T;z_i)$. In Fig.~\ref{fig4} (main panel) the results are shown for a 10\,nm and a 20\,nm thick nano-granular metal layer assuming different orientations of the TTF-CA top layer as indicated in the plot. Before we proceed we note that the parallel circuit is highly simplifying, since it is only directly applicable for diffusive charge transport. Here, due to the disordered nature of the Pt(C) nano-granular layer, a tunnel-percolative transport that minimizes the overall electrostatic energy may be assumed \cite{Middleton1993a}. However, so far no theory was developed that would take both, the percolative nature and the correlated co-tunneling scenario into account. We assume that the effect of charging energy renormalization due to the TTF-CA layer will be qualitatively well described by our approach as long as the thickness of the nano-granular metal layer does not significantly exceed the screening length of about 20\,nm in the present case.

We now briefly describe the main findings from these simulations. For the 20\,nm thick nano-granular metal we find a peak-like anomaly in the temperature-dependent conductance at the PE-FE phase transition of the TTF-CA top layer at 56\,K, if we assume that TTF-CA has its stacking axis in-plane, as in the $c^*$-orientation (blue curve in Fig.~\ref{fig4}, $g=0.01$). For the temperature-dependent dielectric constant of the TTF-CA top layer along the $c^*$-direction we use a Lorentzian shape (see inset of Fig.~\ref{fig4}a, left vertical axis). If, on the other hand, we assume $a$-axis orientation of the TTF-CA layer with a correspondingly larger dielectric constant (see inset of Fig.~\ref{fig4}a, right vertical axis), an overall increase of the temperature-dependent conductance of the nano-granular metal results but no peak structure (not shown). This is a direct consequence of the saturation effect in the renormalization of the charging energy for $\varepsilon_r^{TTF-CA}$ larger than about 100, as already alluded to above. Due to the fact that for $a$-axis orientation $\varepsilon_r^{TTF-CA}$ is already above 50 before and after the phase transition, rising to several 100 at the Curie temperature, the conductance modulation peak at $T_C$ is completely smeared out. This same observation is made assuming a 10\,nm thick nano-granular metal layer. Here we show in Fig.~\ref{fig4}a (red curve) the result of a model calculation assuming $a$-axis orientation of the TTF-CA top layer with a maximum of the dielectric constant of 500 at the Curie temperature. As is apparent, besides an overall damping of the thermally activated behavior, no peak structure is visible in the conductance. We complete the modeling by showing the influence of an increased inter-granular coupling strength. For the 20\,nm thick nano-granular metal the green curve in Fig.~\ref{fig4}a shows the conductance behavior for $g=0.1$ with TTF-CA being in the $c^*$-orientation. Assuming two different inter-granular coupling strength, namely $g=0.01$ and $g=0.1$, we can rather well reproduce the experimentally observed conductance anomalies for the 20\,nm and 100\,nm layer in Fig.~\ref{fig3}. In conjunction with the two different $g$-values, representing the as-grown (100\,nm) and post-growth electron irradiated Pt(C) layer (20\,nm), we assume an increase of the grain diameter from 1.5\,nm to 3.0\,nm, as this is experimentally observed \cite{Porrati2011a}.  Further comparing the simulated conductance behavior with the experimental results we conclude that for the 20\,nm and 100\,nm thick Pt(C) microstrips the TTF-CA top layers are at least partly formed by growth domains with their stacking axis in-plane (such as in $c^*$-orientation) which leads to a conductance peak at $T_C$, whereas for the 10\,nm Pt(C) microstrip the TTF-CA growth domains have solely $a$-axis orientation and thus only show an overall enhanced conductance in the whole temperature range. We do not believe that differing growth preferences exist for TTF-CA on 10\,nm or 20\,nm thick Pt(C) microstrips, but rather attribute this to the accidental orientation of the respective growth domains of the polycrystalline TTF-CA layer covering the respective microstrips.

We conclude this part with a brief remark concerning the noisy character of the data in the phase transition region. Due to the fact that a rather small set of TTF-CA growth domains is covering the Pt(C) microstrip the conductance modulations are not subject to a pronounced averaging effect. As a consequence, the charging energy renormalization acquires a time dependence which stems from charge fluctuations in the TTF-CA FE domains as the domain walls move by thermal activation. Also, if an extended PE-FE coexistence region is also assumed to occur in the clamped TTF-CA layers, any time-dependence in the spatial arrangement of the PE and FE phase volumes will be reflected in the nano-granular metals' conductance. We therefore argue that the Pt(C) microstrips, in particular if further scaled down to the diameter of a typical TTF-CA growth domain, represents a very sensitive and non-invasive probe of the time-dependent dielectric properties of the FE (and PE) layer.

An important observation from our experiments is that the apparent phase transition temperature is shifted by 25\,K from 81\,K to about 56\,K. We would like to present some arguments that this is due to the clamping effect in the following way. From the known temperature-dependent thermal expansion coefficient of Si (bulk of the substrate material) \cite{Ibach1969a} and TTF-CA \cite{Batail1981a} we calculate the thermally-induced clamping strain assuming full clamping as the heterostructure is cooled from the growth temperature (193\,K) to 81\,K. Depending on the orientation of the respective TTF-CA growth domain this leads to strain values of 0.2\,$\%$ to 1.0\,$\%$, with the larger strain value corresponding to an orientation for which the stack axis lies in the substrate plane. The associated stress values can only be estimated from the bulk modulus of TTF-CA \cite{King1986a}, since no measurements of the temperature-dependent elastic constants of TTF-CA are known to us. From this we estimate stress levels of 14\,MPa to 69\,MPa. We now use results for the hydrostatic pressure dependence of the Curie temperature of TTF-CA obtained for single crystals \cite{Lemee1997a}. From these measurements we deduce an almost linear pressure dependence of $dT_C/dP = 0.34$\,K/MPa in the relevant temperature range from 80\,K to about 200\,K. If we use this tentatively for the biaxially clamped TTF-CA thin layers under tensile strain we expect a Curie temperature shift of 5\,K to 23\,K, with the large value for the $c^*$-axis orientation. This analysis is hindered by the non-availability of reference data for the anisotropic elastic constants of TTF-CA and the lack of independent data concerning the influence of biaxial strain on the phase transition temperature. Nevertheless, in conjunction with the model calculation that suggests a predominant $c^*$-axis orientation for the samples which exhibit a pronounced peak at 56\,K in the Pt(C) conductance data, our estimate of the $T_C$ shift of 23\,K is amazingly close to the observed shift. We therefore suppose that it is indeed the clamping effect that is responsible for the reduced Curie temperature. Nevertheless, at this point we would also like to comment on possible implications of size effects on the transition. Among the factors that contribute to the thickness-dependence of the ferroelectric instability in thin films are the depolarizing field, only relevant for perpendicular polarization, and the growing importance of the surfaces and interfaces as the thickness is reduced \cite{Dawber2005a}. As has been discussed before, we attribute the peak-like anomaly in the conductance curves of the Pt(C) microstrips to those regions which are below TTF-CA growth domains with in-plane orientation of the stacking axis, i.e. with in-plane polarization. From this we would assume that depolarization effects are not important. Concerning the influence of the surface and interface we point out that a film thickness of 50\,nm corresponds to almost seventy unit cells of TTF-CA (in stacking direction) which renders unlikely any influence of the film thickness in this range on the ferroelectric properties.
\section{Conclusion}
In this work we suggest a size-scalable heterostructure consisting of a nano-granular metal and a ferroelectric layer in which the nano-granular metal acts as a sensitive and non-invasive probe of the (time-dependent) dielectric properties of the ferroelectric layer. As a consequence of the fabrication technique for the nano-granular metal, focused electron beam induced deposition, the probing structure can be scaled-down to the size of a single growth domain of the ferroelectric layer. With proper design of the probing structure this does also allow for studying the spatial distribution of the dielectric properties. Based on the writing capabilities of FEBID and the typical grain size of 1.5\,nm to 5\,nm for the nano-granular metal, we expect similar performance of the probing structure as described here down to probing areas of 20\,nm to 30\,nm. We have shown this probing effect for the organic FE TTF-CA as a case study and found, as an additional important observation, that clamping can be very efficient in tuning the phase transition temperature for TTF-CA. However, our approach should also be applicable to inorganic ferroelectrics and, in particular, also for a reversed order of the layer structure with the nano-granular metal deposited on the top of a pre-formed, and possibly epitaxial, ferroelectric layer. We conclude by remarking on the observed saturation effect in the charging energy renormalization for dielectric constants significantly above 100. In this case, an even more sensitive probe structure can be envisioned, consisting of a dense array of isolated metallic nano-dot structures between which the ferroelectric (or any other dielectric material) is filled-in as matrix material. Work along these lines is in progress. 
\section{Methods}
As substrate material we used n-doped (100)-oriented Si with a thermally grown amorphous SiO$_2$ layer of 200\,nm thickness. Prior to use the substrates were thoroughly cleaned by acetone, isopropanol and de-ionized water in an ultrasound bath. Au/Cr contacts for the two-probe measurements, as well as interdigitized electrode structures were fabricated by standard photolithography steps. Atomic force microscopy data were acquired in non-contact mode employing a nanosurf easyscan 2 microscope. The X-ray diffraction experiments were performed on a Bruker thin film diffractometer with parallel beam optics employing a Cu anode. The temperature-dependent conductance data were taken at a constant bias voltage (Keithley sourcemeter 2400 and 2636A) in a variable temperature insert inside a $^4$He cryostat operating in the range from 1.4\,K to 300\,K. For the conductance measurement on the thick TTF-CA film on interdigitized electrodes the bias voltage was set to 10\,V corresponding to an electric field between neighboring electrodes of $3\times 10^4$\,V/cm in order to get sufficiently large currents. Conductance measurements on the Pt(C) microstrips were taken with electric field values of typically 100\,V/cm to 1000\,V/cm. The dielectric measurements were performed at 10\,V ($3.3\times 10^4$\,V/cm) and 111\,Hz employing an LCR meter (HP 4284A).

The nano-granular microstrip structures have been prepared by focused electron beam induced deposition using the precursor Me$_3$CpMePt(IV) (Cp: cyclopentadienyl, Me: methyl) in an adapted dual-beam scanning ion/electron microscope with Schottky emitter operating a 5\,kV beam voltage and 1.6\,nA beam current. The tuning of the inter-granular tunnel coupling strength has been accomplished by a post-growth electron irradiation process with a dose of 100\,nC/$\mu$m$^2$ in the present case. Details of the fabrication and inter-granular tunnel coupling tuning process can be found in \cite{Porrati2011a}.

For the TTF-CA thin film growth we employed a custom-build molecular beam deposition system with a base pressure of $5\times 10^{-7}$\,mbar. The TTF-CA thin films were grown by single-source evaporation of stoichiometrically mixed proportions of TTF and CA crystallites at 366\,K effusion cell temperature using a quartz liner. The growth rate was set to approximately 0.8\,nm/s for the 50\,nm layers. The source materials were used as supplied (Alfa Aesar, purity $\geq 97\,\%$). In several preparatory experiments we found that the sticking coefficients of TTF and CA for different substrate materials (SiO$_2$, NaCl (100), sapphire) held at room temperature is extremely low, so that stable thin film growth conditions could not be achieved. In stepwise reduction of the substrate temperature down to about 190\,K with liquid nitrogen cooling we determined the temperature-dependent sticking coefficients of TTF and CA and set the substrate temperature for the TTF-CA thin film growth to 193\,K. A more detailed account of the growth characteristics of TTF-CA thin films will be given in a separate publication.
\fig{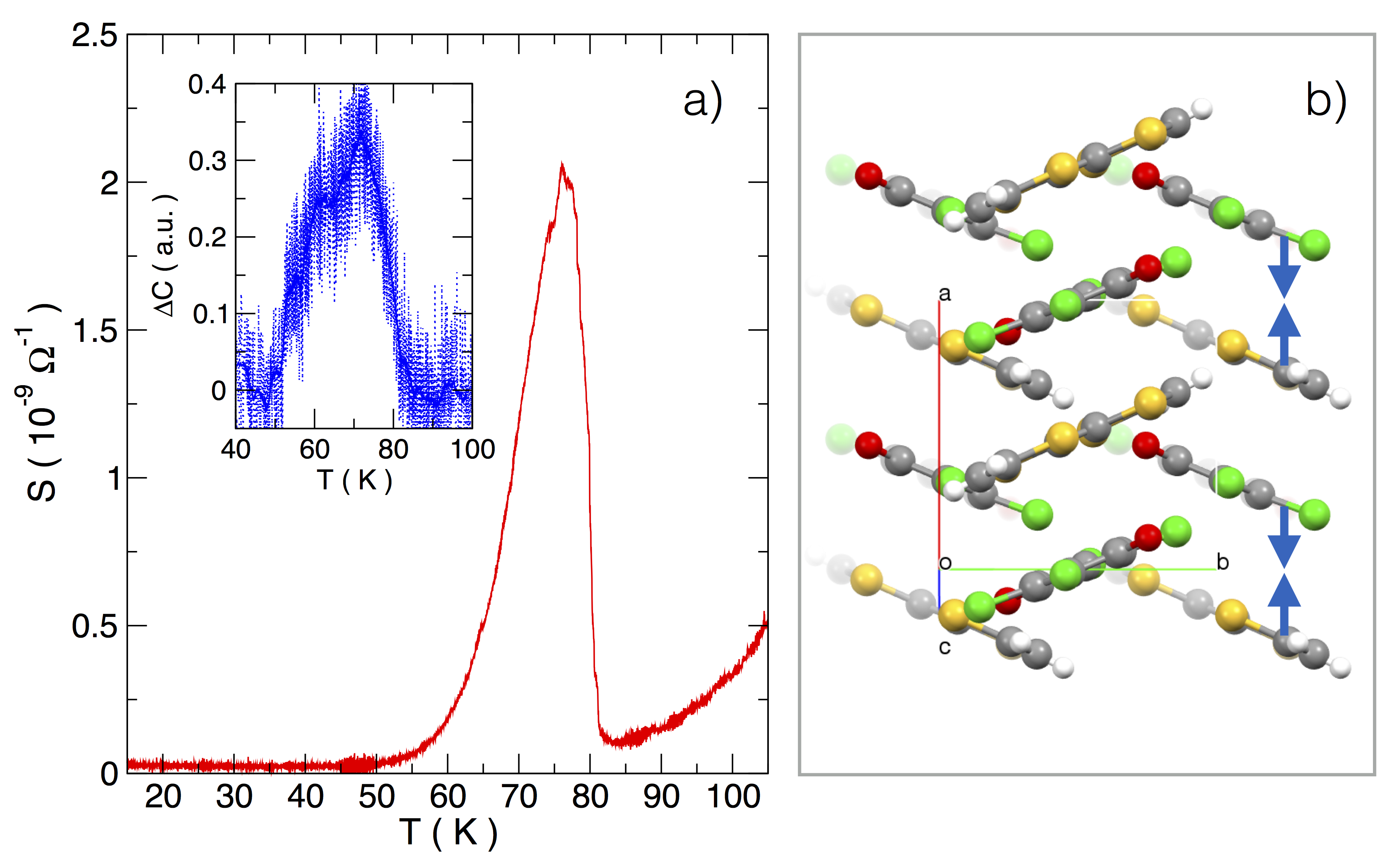}{0.9}{{\bf Selected properties of TTF-CA.} (a) Temperature-dependent conductance of micrometer thick TTF-CA layer deposited on Si/SiO$_2$ substrate with photolithographically pre-defined interdigitized electrode structure. The PE-FE phase transition shows the typical conductance increase of bulk samples starting at 84\,K and extending down to about 50\,K. Below 50\,K the sample resistance exceeds the isolation resistance of our setup (about 30\,G$\Omega$). Inset: Peak in the interdigitized electrode structure capacitance caused by PE-FE transition of the TTF-CA layer. The dotted line represents the data as-measured, whereas the blue solid line depicts a running average of 20 data points. (b) Crystal structure of TTF-CA viewed along the $c^*$-axis. The TTF molecules can easily be discriminated by the S atoms (yellow coloring), whereas the CA molecules can be identified by the Cl atoms (green coloring). The blue arrows indicate the direction of the dimerization of D/A-pairs at the PE-FE phase transition. }{fig1}
\fig{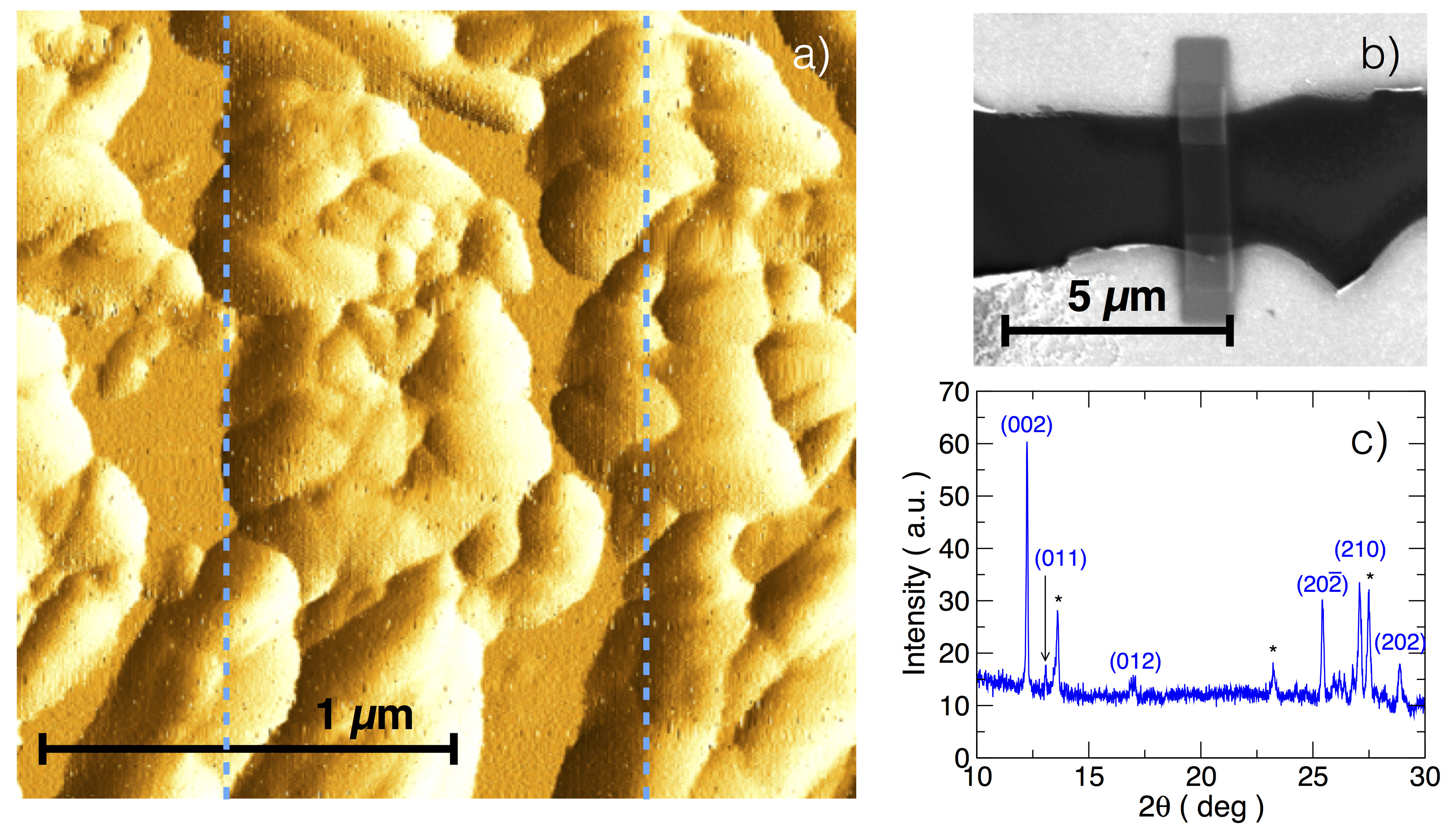}{0.9}{{\bf Microstructure details of TTF-CA thin films on Pt(C).} (a) AFM image of nominally 50\,nm thick TTF-CA layer on Si/SiO$_2$ showing TTF-CA islands of 500\,nm to 1\,$\mu$m diameter leaving a small fraction of the substrate surface free. The blue dashed line indicates the width of a typical Pt(C) microstrip on which the TTF-CA is deposited. (b) SEM image of 20\,nm thick Pt(C) microstrip of $5\times 1$\,$\mu$m$^2$ lateral dimension between Au contact structures. At the positions where the thin Pt(C) microstrip crosses over to the Au contacts thicker Pt(C) patches are used to ensure good electrical contact to the electrodes. (c) X-ray Bragg scan of typical 50\,nm thick TTF-CA layer on Si/SiO$_2$ substrate. The polycrystalline microstructure is apparent from the reflection indices. The asterisks mark an impurity phase which we identified as the insulating black polymorph of TTF-CA \cite{Benjamin2011a}.}{fig2}
\fig{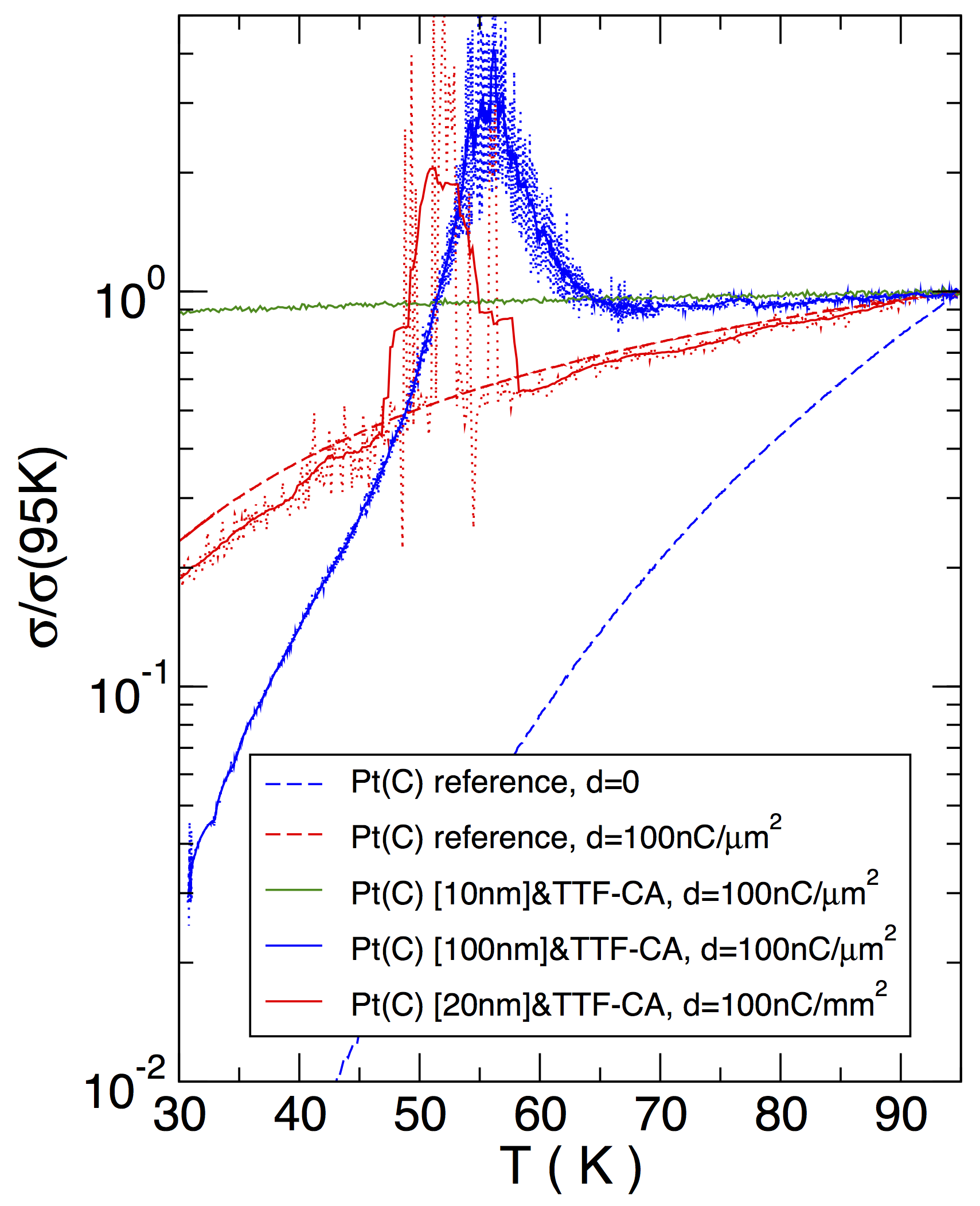}{0.6}{{\bf Conductance modulation in Pt(C) by TTF-CA top layers.} Temperature-dependent conductance data of Pt(C) microstrips of different thickness with TTF-CA top layer. $d$ denotes the respective doses used in a post-growth electron irradiation step employed to increase the inter-granular tunnel coupling strength $g$ of the granular metal (see \cite{Sachser2011a} for details). The associated conductance increase of the Pt(C) microstrips by post-growth irradiation becomes apparent by comparison with the reference curve (blue dashed line). The data is normalized to the respective conductance value of the Pt(C) microstrips at 95\,K for clarity. Due to their high resistance values the TTF-CA top layers do not contribute to the measured conductance. The dashed lines (red and blue) represent the results obtained on uncovered Pt(C) microstrips. The dotted lines show the data for TTF-CA-covered Pt(C) microstrips as measured, whereas the solid lines are obtained from a running average over 20 data points.}{fig3}
\fig{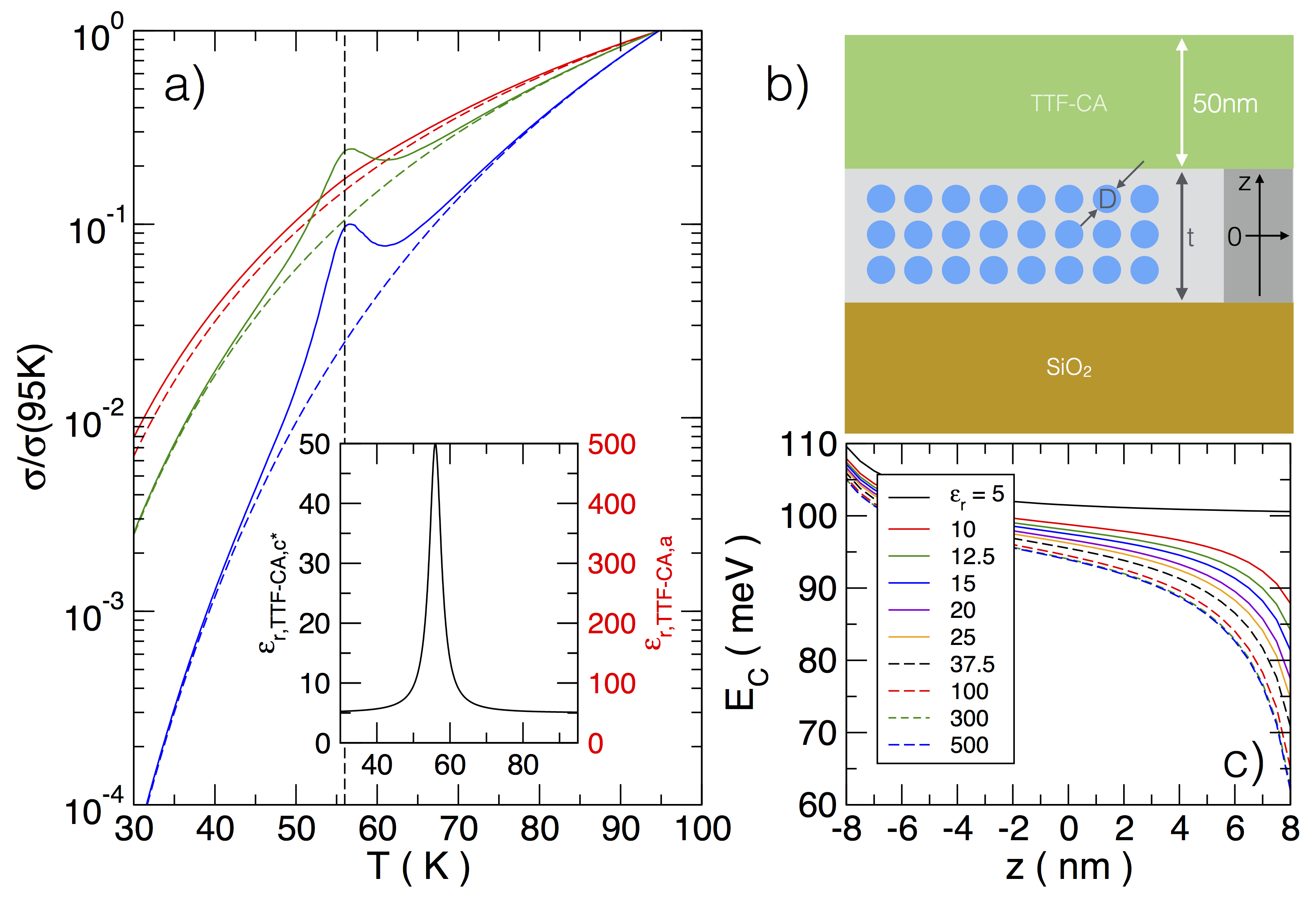}{0.9}{{\bf Model calculations for Pt(C) conductance modulation by TTF-CA.} (a) Result of model calculation of the temperature-dependent conductance of a nano-granular metal within a three-layer structure as schematically depicted in (b). The temperature dependence of the dielectric constant of the TTF-CA top layer in the two main axis directions $a$ (right y-axis in inset) and $c^*$-axis (left y-axis in inset) is assumed to follow a Lorentzian with a half-width at half maximum of 2\,K. The dashed line indicates the Curie temperature set to 56\,K. Blue line: Pt(C) thickness is 20\,nm, Pt grain size is 1.5\,nm, coupling strength is $g=0.01$. Green line: Pt(C) thickness is 20\,nm, Pt grain size is 3.0\,nm, coupling strength is $g=0.1$. Red line: Pt(C) thickness is 10\,nm, Pt grain size is 3.0\,nm, coupling strength is $g=0.1$. The dashed line represent the calculated Pt(C) conductance without TTF-CA top layer. See text for details. (c) Set of charging energies for spherical Pt particle of 3\,nm diameter at different positions below the TTF-CA top layer as indicated. The different curves were calculated for different $\varepsilon_r^{TTF-CA}$ as indicated. For all simulations the dielectric constant of the Pt(C) layer as effective medium was set to 5.0 and for SiO$_2$ it was set to 2.8.}{fig4}
\end{document}